\documentclass[]{aa}
\usepackage{graphicx,pifont,amssymb}
\begin{document}
\title{Multiple protostellar systems. I. A deep near infrared survey of
  Taurus and Ophiuchus protostellar objects\thanks{Based on
  observations obtained at the Canada-France-Hawaii Telescope
  and at the European Southern Observatory, La Silla, Chile.}}

   \author{G. Duch\^ene\inst{1,2}
          \and
          J. Bouvier\inst{2}
          \and
          S. Bontemps\inst{3}
          \and
          P. Andr\'e\inst{4}
          \and
          F. Motte\inst{4}
          }

   \offprints{G. Duch\^ene, \email{gaspard.duchene@obs.ujf-grenoble.fr}}

   \institute{Department of Physics \& Astronomy, UCLA, Los
Angeles, CA 90095-1562, USA
    \and Laboratoire d'Astrophysique de Grenoble, Universit\'e
Joseph Fourier, BP 53, 38041 Grenoble Cedex 9, France
    \and Observatoire de Bordeaux, BP 89, 33270 Floirac, France 
    \and CEA/DSM/DAPNIA, Service d'Astrophysique, CEA Saclay,
91191 Gif-sur-Yvette Cedex, France }

\authorrunning{Duch\^ene et al.}
\titlerunning{Multiple protostellar systems}

   \date{Received 3 May 2004; accepted 22 July 2004}
   
   \abstract{We performed a deep infrared imaging survey of 63
   embedded young stellar objects (YSOs) located in the Taurus
   and Ophiuchus clouds to search for companions. The sample
   includes Class\,I and flat infrared spectrum protostellar
   objects. We find 17 companions physically bound to 15 YSOs
   with angular separations in the range 0.8-10\arcsec
   (110--1400\,AU) and derive a companion star fraction of
   23$\pm$9\% and 29$\pm$7\% for embedded YSOs in Taurus and
   Ophiuchus, respectively, about twice as large as that found
   among G dwarfs in the solar neighborhood. Therefore, binary
   and multiple protostellar systems are a very frequent
   outcome of the fragmentation of prestellar cores. In spite
   of different properties of the clouds and especially of the
   prestellar cores, the fraction of wide companions,
   27$\pm$6\,\% for the combined sample, is identical in the
   two star-forming regions. This suggests that the frequency
   and properties of wide multiple protostellar systems are
   not very sensitive to specific initial
   conditions. Comparing the companion star fraction of the
   youngest YSOs still surrounded by extended envelopes to
   that of more evolved YSOs, we find evidence for a possible
   evolution of the fraction of wide multiple systems, which
   seems to decrease by a factor of about 2 on a timescale of
   $\sim$10$^5$\,yr. For the first time, it is possible to
   confront the result of a multiplicity survey of a nearly
   complete population of embedded YSOs at an age of
   $\sim$10$^5$\,yr to numerical simulations of molecular
   cloud collapse which, after a few free fall times, reach
   this evolutionary stage.  Somewhat contrary to model
   predictions, we do not find evidence for a sub-clustering
   of embedded sources at this stage on a scale of a few
   100\,AU that could be related to the formation of small-N
   protostellar clusters. Possible interpretations for this
   discrepancy are discussed.  \keywords{Stars: formation --
   (Stars:) binaries: visual -- Stars: pre-main sequence -- }
   }

   \maketitle
%

\section{Introduction}
\label{sec:intro}

Giant molecular clouds produce stars in a multi-step process
that needs to be understood in order to account for the
general properties of stars, such as the stellar mass function
or the properties of multiple systems in the Galaxy. First,
the molecular cloud fragments into individual cores that
subsequently collapse to form individual stars. In many cases,
however, those cores fragment a second time during
protostellar collapse (in the following, we refer to this
phenomenon as ``core dynamical fragmentation''), which leads
to the formation of binary and higher order multiple
systems. The extent to which this process depends upon initial
physical conditions of the cloud is currently under
debate. Numerical simulations suggest that the dynamical
fragmentation process is strongly dependent on the initial
conditions, such as cloud rotation, gas temperature or
magnetic field strength (e.g., Bonnell et al. 1992; Durisen \&
Sterzik 1994; Boss 2002).  With several star-forming regions
located within 500\,pc of the Sun that range from relatively
quiescent zones such as the Taurus-Auriga cloud to the dense
cluster-forming Orion Nebula, a number of observational tests
can be performed to search for environment-induced effects.

The study of the collapse and core fragmentation processes
usually focuses on two main approaches. Investigation of
prestellar cores within molecular clouds provides insight on
the initial conditions and the very early phases of the
collapse while the analysis of populations of pre-main
sequence (T\,Tauri) objects reveal their outcome. Due to the
limited spatial resolution of the single-dish millimeter
images from which prestellar core properties are derived, it
is usually not possible to probe spatial scales of a few tens
or hundreds AU, typical of most multiple systems. On the other
hand, T\,Tauri stars have not necessarily preserved their
original properties after the dissipation of their collapsing
envelope. Studying more deeply embedded young stellar objects
(YSOs), which are assumed to be much younger than T\,Tauri
stars, at high spatial resolution ($\leq1$\arcsec) offers an
opportunity to sample a crucial intermediate stage between
prestellar cores and pre-main sequence stars.

Recent studies have revealed ranges in the properties of prestellar
cores as well as in the young populations resulting from the star
formation process, with significant differences between star-forming
regions. We summarize here some of the most notable ones, with
emphasis on the Taurus and Ophiuchus molecular clouds which constitute
the focus of the present study. First of all, the Taurus molecular
cloud forms only a few tens of objects in $\sim1$\,pc-radius groups
(Gomez et al. 1993) and is frequently referred to as the prototype of
``isolated'' star formation. It is believed that most stars form in
much richer, cluster-like structures with several hundred objects
within a 1\,pc-radius (e.g., Carpenter 2000; Adams \& Myers 2001).
While the Orion Trapezium cluster is a well known example of this
``clustered'' mode of star formation, the Ophiuchus molecular cloud
also belongs to this category, as it host several hundred objects with
an average spatial density on the order of $10^3$\,stars/pc$^3$
(Bontemps et al.  2001; Allen et al. 2002).  Taurus and Ophiuchus are
therefore excellent candidates to investigate whether the star
formation process is sensitive to environmental conditions.

Millimeter continuum mapping of both star-forming regions have
revealed clear differences in the properties of their prestellar
cores. While cores in Taurus appear to have a radius as large as
10000--20000\,AU and smoothly merge into the ambient molecular cloud
(Motte \& Andr\'e 2001), those found in Ophiuchus exhibit a sharp
boundary at a radius on the order of 5000\,AU (Motte, Andr\'e \& Neri
1998), possibly due to external perturbations in this denser
environment. The derived envelope masses, bolometric luminosities and
density profiles of prestellar cores also differ between the two
molecular clouds. Likewise, at a slightly later stage, the
circumstellar envelopes of Ophiuchus embedded protostars are smaller
than those of their Taurus counterparts (Motte et al. 1998). This led
Motte \& Andr\'e (2001) to state that ``the beginning of protostellar
evolution is suggested to be more violent (\dots ) in protoclusters
compared to regions of distributed star formation like Taurus''. In
other words, the initial conditions for core fragmentation appear to
be significantly different in the Taurus and Ophiuchus molecular
clouds.

Multiple systems are useful tracers of the core fragmentation
process.  Systematic surveys have revealed that young stellar
populations in dense stellar clusters (Orion Trapezium,
IC\,348, Pleiades) exhibit multiplicity rates similar to field
stars whereas populations of T\,Tauri stars located in more
distributed associations can possess up to twice as many
companions (e.g., Duch\^ene 1999). However, it must be noted
that the frequency of visual companions to T Tauri stars in
the Taurus and Ophiuchus clouds are identical, both exhibiting
a strong excess of multiple systems, i.e., that the latter
behaves in the same way as ``loose associations'' in this
respect although it is an example of clustered star formation.

The lower multiplicity rate in dense clusters does not
necessarily imply that the core fragmentation process itself
is environment-dependent. For instance, Kroupa, Petr \&
McCaughrean (1999) have performed N-body simulations to follow
the dynamical evolution of young stellar clusters and showed
that the observed properties of multiple systems in these
populations are consistent with an {\it initial} set of
multiple systems identical to loose T\,Tauri associations. In
clusters, direct dynamical encounters between systems are able
to disrupt a fraction of the initial (wide) systems over
timescales as short a 1\,Myr (e.g., Kroupa 1995a), thus
rapidly bringing the multiplicity rate down to the observed
value.

In order to probe the outcome of the fragmentation process
before it is dynamically modified, at least in clustered
environments, it is therefore necessary to study younger YSOs
than T\,Tauri stars. In this study, we present a systematic
search for close companions among Class\,I sources and
slightly more evolved ``flat-spectrum'' objects in
Taurus-Auriga and Ophiuchus. These objects exhibit a flat or
rising spectral energy distribution from the near to the mid
infrared and are believed to still be heavily embedded within
a collapsing envelope. On a statistical basis, their estimated
age is of order of $10^5$\,yrs (e.g. Kenyon \& Hartmann 1995).
Because they are seen through a high column density of dust,
these objects are almost undetected in the visible and deep
near infrared imaging is required to look for close
companions. Recently, Haisch et al. (2004) reported the
results of an imaging survey of samples of embedded YSOs
located in several nearby molecular clouds and derived a
multiplicity rate over a separation range 300-2000\,AU similar
to that of more evolved T\,Tauri stars but they did not
conclude regarding possible differences between various
star-forming regions. Our survey differs from theirs in two
respects. Firstly, we focus on homogeneous and nearly complete
samples of embedded YSOs in two star-forming regions only,
Taurus and Ophiuchus, in order to be able to contrast at a
significant statistical level the multiplicity rate resulting
from the isolated and clustered star formation modes. Second,
the 0.6\arcsec\ angular resolution of the present survey
allows us to explore closer systems down to a separation of
110\,AU at the distance of these two clouds (140 pc, Bertout,
Robichon \& Arenou 1999; Bontemps et al. 2001).

\begin{table*}
\caption{\label{tab:sample1}Sample of protostars surveyed 
in Taurus. The infrared photometry and the spectral indices
are taken from Kenyon \& Hartmann (1995), except for
IRAS\,04191+1523, whose $JHK$ photometry is from the 2MASS
database and classification from Motte \& Andr\'e (2001). The
bolometric luminosities are from Motte \& Andr\'e (2001). IRAS
did not resolve the systems IRAS\,04108+2803\,AB and
IRAS\,04181+2654\,AB and we assigned the same spectral index,
based on unresolved photometry, to both components and split
evenly the bolometric luminosity of the later system. The
classification of each source is based on the infrared
spectral indices (see text); ``I'' stands for Class\,I sources
and ``FS'' for flat spectrum sources. The last three columns
indicate whether the source i) has at least one visual
companion within 10\arcsec, ii) is surrounded by an extended
nebulosity in our images (see also Park \& Kenyon 2002), and
iii) has an extended ($>1000$\,AU) envelope as seen in
millimeter continuum mapping (from Motte \& Andr\'e 2001).}
\begin{tabular}{lcccccccccc}
\hline
Object & $J$ & $H$ & $K$ & $\alpha^{2-12}_{IR}$ &
$\alpha^{2-25}_{IR}$ & Class & $L_{bol} (L_\odot)$ & Comp? &
Opt neb? & Mm env? \\
\hline
IRAS\,04016+2610 & 13.62 & 11.67 & 9.33 & 1.06 & 1.02 & I &
3.7 & N & Y & N \\
IRAS\,04108+2803\,B & 16.25 & 13.25 & 11.12 & 0.50 & 0.62 & FS
& 0.6 & N & N & N \\
IRAS\,04113+2758 & 11.29 & 9.15 & 7.79 & -0.01 & 0.12 & FS &
$>1.6$ & Y & N & Y \\
IRAS\,04158+2805 & 13.15 & 11.92 & 10.80 & 1.47 & 0.72 & I &
$>0.4$ & N & Y & N \\
IRAS\,04169+2702 & 16.87 & 13.83 & 11.22 & 0.89 & 1.10 & I &
0.8 & N & Y & Y \\
IRAS\,04181+2655 & 15.46 & 12.74 & 10.54 & 0.60 & 0.55 & I &
0.4 & N & N & N \\
IRAS\,04181+2654\,A & 15.90 & 12.74 & 10.58 & -0.00 & 0.10 &
FS & 0.25 & N & N & Y \\
IRAS\,04181+2654\,B & 18.11 & 13.64 & 10.81 & -0.00 & 0.10 &
FS & 0.25 & N & N & Y \\
IRAS\,04191+1523 & 16.74 & 14.43 & 12.26 & & 
& I & 0.5 & Y & Y & Y \\
IRAS\,04239+2436 & 15.61 & 12.96 & 10.58 & 1.27 & 1.15 & I &
1.3 & N & N & Y \\
IRAS\,04248+2612 & 12.76 & 11.39 & 10.65 & -0.09 & 0.48 & FS &
0.4 & Y & Y & Y \\
IRAS\,04260+2642 & 14.64 & 12.96 & 11.63 & 0.16 & 0.25 & FS &
0.1 & N & N & N \\
IRAS\,04263+2426 & 10.84 & 9.06 & 7.80 & 1.02 & 0.74 & I & 7.0
& Y & Y & Y \\
IRAS\,04264+2433 & 13.63 & 12.07 & 11.09 & 0.67 & 0.98 & I &
0.5 & N & N & N \\
IRAS\,04295+2251 & 13.62 & 11.11 & 9.55 & 0.11 & 0.21 & FS &
0.6 & N & N & N \\
IRAS\,04302+2247 & 14.20 & 12.35 & 11.07 &  & 0.15 &
FS & 0.3 & N & Y & N \\
IRAS\,04325+2402 &  & 13.02 & 11.03 &  & 0.79
& I & 0.9 & Y & Y & Y \\
IRAS\,04361+2547 & 16.07 & 12.88 & 10.55 & 1.27 & 1.55 & I &
3.7 & N & Y & Y \\
IRAS\,04365+2535 & 17.28 & 13.39 & 10.62 & 1.08 & 1.27 & I &
2.4 & N & Y & Y \\
IRAS\,04381+2540 & 16.21 & 14.32 & 12.00 & 1.20 & 1.31 & I &
0.7 & N & Y & Y \\
IRAS\,04385+2550 & 12.61 & 10.72 & 9.65 & 0.14 & 0.18 & FS &
$>0.4$ & N & N & N \\
IRAS\,04489+3042 & 13.74 & 11.58 & 10.11 & 0.18 & 0.31 & FS &
0.3 & N & N & N \\
\hline
\end{tabular}
\end{table*}

This paper is organized as follows: in \S\,\ref{sec:obs}, we
present our samples and deep infrared observations obtained at
CFHT and ESO/NTT.  Candidate visual companions detected on the
images are presented in \S\,\ref{sec:res} and physical
companions to embedded YSOs are identified on statistical
grounds, thus allowing us to establish the multiplicity
frequency of embedded YSOs in the Taurus and Ophiuchus
clouds. In \S\,\ref{sec:discuss}, we discuss the results and
their implications regarding the fragmentation of prestellar
cores in the two star-forming regions, and compare the
multiplicity frequency we derive for embedded Class I and flat
spectrum sources to that of both younger Class\,0 sources and
older Class\,II-III T Tauri stars in the same clouds in order
to investigate the dynamical evolution of protostellar systems
on a timescale of a few Myr.  Finally, we summarize our
results in \S\,\ref{sec:concl}.


\section{Sample selection and observations}
\label{sec:obs}

\begin{figure}
\centering
\includegraphics[width=\columnwidth]{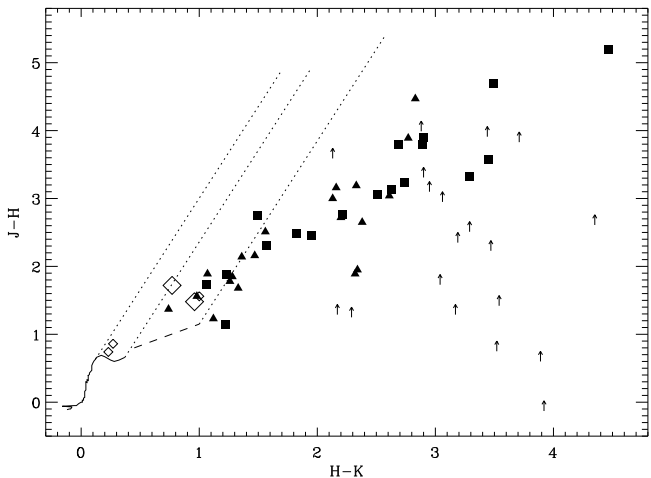}
\caption{Near-infrared color-color magnitude for all our
targets in Taurus ({\it filled triangles}) and Ophiuchus ({\it
filled squares}). Objects not detected by 2MASS in the $J$
band are indicated with arrows. The companions detected at all
three wavelengths in our survey are indicated by diamonds, the
larger ones representing those that are likely to be bound to
one of our targets. The solid curve indicates the locus of
main sequence sources and the dashed line represents the
dereddened locus of all T\,Tauri stars in the Taurus region
(Meyer 1996). The dotted lines delineate the reddening band of
both loci for $A_V=25$\,mag and the extinction law from Rieke
\& Lebofsky (1985).}
\label{fig:jhk}
\end{figure}

We have built our samples from the object lists from Motte \&
Andr\'e (2001) in Taurus and Bontemps et al. (2001) in
Ophiuchus. The list of our targets is presented in
Tables\,\ref{tab:sample1} and \ref{tab:sample2}, together with
their near-infrared photometry and near- to mid-infrared
spectral index.  The latter is used to distinguish between
Class\,I, flat-spectrum and T\,Tauri (Class\,II) sources,
where we adopted $\alpha_{IR}=0.55$ and $\alpha_{IR}=-0.05$
thresholds between these categories (see Bontemps et
al. 2001). Note that because of the large line-of-sight
extinctions towards some of our targets, the use of the
observed infrared spectral index is not completely
secure. While there might be a few embedded T\,Tauri stars in
our sample, especially among the objects classified as flat
spectrum, we consistently use the observed spectral index to
estimate each object's nature since we cannot estimate the
extinction to each individual source.  Overall, we have
observed all known sources in both areas with the exception of
two Class\,I sources and one flat-spectrum object in Taurus,
all of which are in the crowded vicinity of HL\,Tau, to the
south of the molecular cloud. Our sample therefore includes 22
protostellar sources in Taurus and 41 in Ophiuchus, in the
magnitude range $7.5\leq K\leq 14.5$ with a median of
$K\sim11$\,mag. A near-infrared color-color diagram of all our
targets is presented in Figure~\ref{fig:jhk}, which shows that
most sources have redder colors than T\,Tauri stars, even if
extinction is taken into account.

\begin{table*}
\caption{\label{tab:sample2}Sample of protostars surveyed in
Ophiuchus. ISO numbers, infrared spectral indices and bolometric
luminosities are from Bontemps et al. (2001), while the near-infrared
magnitudes are from Barsony et al. (1997), Greene et al. (1994) and
the 2MASS database (2MASS uses a $Ks$ filter rather than a $K$ filter;
the difference in magnitude for an object can reach a few tenths of a
magnitude). The source marked with a $\dagger$ symbol has an infrared
spectral index typical of a normal Class\,II source but Ressler \&
Barsony (2001) have shown that this system contains a Class\,I source
and it is classified as such here. The last three columns have the
same meaning as in Table~\ref{tab:sample1}. The existence of an
envelope is taken either from Andr\'e \& Montmerle (1994) and Motte et
al. (1998) or from our reexamination of the mosaic presented in the
latter. (Note that this classification is more uncertain than in
Taurus due to the smaller size of protostellar envelopes in
Ophiuchus. Particularly uncertain cases are flagged with a ``?''.)}
\begin{tabular}{clccccccccc}
\hline
ISO \# & Object & $J$ & $H$ & $K$ & $\alpha^{2-14}_{IR}$ &
Class & $L_{bol} (L_\odot)$ & Comp? & Opt neb? & Mm env? \\
\hline
21 & CRBR\,12 & $>$17.00 & 15.58 & 12.04 & 0.91 & I & 0.42 & N
& N & N \\
26 & CRBR\,15 & 16.35 & 13.89 & 11.94 & 0.01 & FS & 0.083 & N
& N & N \\
29 & GSS\,30, GY\,6 & 13.89 & 10.83 & 8.32 & 1.20 & I & 21.& N
& Y & Y? \\
31 & LFAM\,1 & $>$17.00 & $>$15.50 & 13.59 & 1.08 & I & 0.13 &
Y & N & Y \\
33 & GY\,11 & 16.52 & 15.37 & 14.15 & 0.31 & FS & 0.011 & Y &
Y & N \\
46 & VSSG\,27, GY\,51 & 16.69 & 13.46 & 10.72 & 0.17 & FS &
0.41 & Y & N & N \\
54 & GY\,91, CRBR\,42 & $>$17.00 & 16.40 & 12.51 & 0.70 & I &
0.17 & N & N & Y? \\
65 & WL\,12, GY\,111 & 16.86 & 13.07 & 10.18 & 1.04 & I & 2.6
& N & Y & Y? \\
70 & WL\,2, GY\,128 & $>$17.00 & 14.05 & 10.99 & 0.05 & FS &
1.4 & Y & N & N \\
75 & GY\,144 & $>$17.00 & 15.75 & 13.46 & 0.20 & FS & 0.053 &
N & N & N \\
77 & GY\,152 & $>$17.00 & $>$15.75 & 13.64 & 0.05 & FS & 0.052
& N & N & N? \\
85 & CRBR\,51 & $>$17.00 & $>$15.75 & 14.00 & 0.03 & FS &
0.035 & N & N & N? \\
99 & LFAM\,26, GY\,197 & $>$17.00 & $>$17.90 & 14.63 & 1.25 &
I & 0.064 & N & Y & Y \\
103 & WL\,17, GY\,205 & 16.90 & 13.57 & 10.28 & 0.42 & FS &
0.88 & N & N & Y? \\
108 & EL\,29, GY\,214 & 17.21 & 12.01 & 7.54 & 0.98 & I &
26. & N & Y & Y \\
112 & GY\,224 & $>$17.00 & 13.69 & 10.79 & 0.34 & FS & 1.7 & N
& N & N? \\
118 & IRS\,33, GY\,236 & $>$17.00 & 15.27 & 12.23 & 0.32 & FS
& 0.43 & N & N & N? \\
119 & IRS\,35, GY\,238 & $>$17.00 & 16.25 & 12.73 & 0.30 & FS
& 0.45 & N & N & N \\
121 & WL\,20, GY\,240 $\dagger$ & 13.45 & 10.78 & 9.21 & -0.07
& I & 1.5 & Y & N & Y? \\
124 & IRS\,37, GY\,244 & $>$17.00 & 13.90 & 10.95 & 0.35 & FS
& 1.5 & Y & Y & N \\
127 & GY\,245 & 19.00 & 15.43 & 11.98 & 0.17 & FS & 0.13 & N &
N & N \\
129 & WL\,3, GY\,249 & $>$17.00 & 14.49 & 11.20 & 0.23 & FS &
1.6 & N & Y & N? \\
132 & IRS\,42, GY\,252 & 15.21 & 11.31 & 8.41 & 0.08 & FS &
5.6 & N & N & N? \\
134 & WL\,6, GY\,254 & $>$17.00 & 14.39 & 10.04 & 0.59 & I &
1.7 & N & N & N \\
137 & CRBR\,85 & $>$17.00 & 17.13 & 13.21 & 1.48 & I & 0.36 &
N & N & Y \\
139 & GY\,260 & $>$17.00 & 15.71 & 12.54 & -0.03 & FS & 0.30 &
N & N & N \\
141 & IRS\,43, GY\,265 & $>$17.00 & 13.17 & 9.46 & 0.98 & I &
6.7 & Y & Y & Y \\
143 & IRS\,44, GY\,269 & $>$17.00 & 13.09 & 9.65 & 1.57 & I &
8.7 & N & Y & Y \\
145 & IRS\,46, GY\,274 & $>$17.00 & 14.65 & 11.46 & 0.94 & I &
0.62 & N & N & Y? \\
147 & IRS\,47, GY\,279 & 15.44 & 11.64 & 8.95 & 0.17 & FS &
3.7 & N & Y & N \\
159 & IRS\,48, GY\,304 & 10.53 & 8.65 & 7.42 & 0.18 & FS & 7.4
& N & N & N \\
161 & GY\,301 & $>$17.00 & 14.77 & 11.30 & 0.12 & FS & 1.8 & N
& N & N? \\
165 & GY\,312 & 16.23 & 13.75 & 11.93 & 0.03 & FS & 0.091 & N
& N & N \\
167 & IRS\,51, GY\,315 & 17.12 & 12.42 & 8.93 & -0.04 & FS &
1.1 & Y & N & Y? \\
170 & B\,162741-244645 & $>$17.00 & 15.71 & 13.54 & 0.51 & FS
& 0.065 & N & N & N \\
175 & GY\,344 & 16.81 & 14.05 & 11.84 & 0.02 & FS & 0.10 & N &
Y & N \\
182 & IRS\,54, GY\,378 & 16.63 & 13.50 & 10.87 & 1.76 & I &
6.6 & Y & Y & Y? \\
200 & ISO\,1631434-245524 & 14.31 & 12.00 & 10.43 & 0.22 & FS
& 1.3 & N & N & N \\
203 & ISO\,1631520-245536 & $>$18.45 & 14.86 & 12.73 & 1.19 &
I & 1.0 & N & Y & N? \\
209 & IRS\,67 & $>$17.30 & 13.31 & 10.43 & 0.75 & I & 1.5 & Y
& Y & Y \\
210 & ISO\,1632021-245616 & 15.81 & 14.08 & 13.02 & 0.09 & FS
& 0.094 & Y & N & N \\
\hline
\end{tabular}
\end{table*}

We conducted most of our observations with the wide-field
infrared camera CFHT-IR (Starr et al. 2000) at
Canada-France-Hawaii Telescope, a 1024$\times$1024 detector
which offers a pixel scale of 0\farcs211/pixel and a total
field-of-view of 3\farcm6. The Ophiuchus images were obtained
on June, 11th and 12th 2001, while the Taurus-Auriga dataset
was obtained on October, 29th and 30th 2001. Each target was
systematically surveyed in the $K$ band, where these
protostars are brighter. In some multiple systems, we obtained
follow-up $H$ and $J$ images to compare the near-infrared
colors of the primary to its candidate companions. For each
field, a sequence of 4 images was obtained, each of them being
cubes of 4 exposures. Individual integration times ranged from
0.1\,s to 40\,s at $K$ and up to 1\,min at $H$ and $J$. In
some Ophiuchus fields, we obtained two sets of images at the
same location, with a short and a long integration time in
order to observe with similar contrasts both bright and faint
targets in crowded fields. The median value and standard
deviation of the full width at half maximum (FWHM) measured on
a sample of single stars throughout the runs are
0\farcs60$\pm$0\farcs07 and 0\farcs65$\pm$0\farcs10 for the
Ophiuchus and Taurus datasets respectively.

On March, 10th 2000, we obtained additional images of the
L\,1689 area, a core located to the south of the main cores of
the Ophiuchus molecular cloud, with ESO's New Technology
Telescope at la Silla Observatory. We used SOFI, which has a
1024$\times$1024 detector, a 0\farcs288/pixel scale and
4\farcm9 field-of-view centered on the known T\,Tauri star
L\,1689-IRS\,5.  With all three $J$, $H$ and $K$ filters, we
obtained a sequence of 10$\times$1.18\,s exposures using a
random jittering pattern in a 20\arcsec$\times$20\arcsec\
box. The final images contain two Class\,I sources and two
flat spectrum objects. The image quality was comparable to
that of our CFHT images, with a 0\farcs55 FWHM, which enables
us to combine this image to our survey.

All images of a given object with the same integration time
were median-combined to obtain a sky estimate, that was then
subtracted from each individual image. The images were then
flat-fielded, aligned based on the location of the brightest
unsaturated source in the field, and median combined to
produce the final images used in this study. The astrometry of
the multiple systems was estimated using the known pixel scale
and orientation of the detectors; uncertainties are on the
order of 0.5\,\% for the plate scale and 1\degr\ for the
orientation.


\section{Results}
\label{sec:res}

The objective of our study is to identify physical companions
to embedded protostars in Taurus and Ophiuchus. We first
identify all visual companions in the 0.8-10\arcsec\
(110--1400\,AU) separation range (\S\,\ref{subsec:comps}) and
compare our results to those of previous studies of protostars
in both areas (\S\,\ref{subsec:compar}).  We then estimate
which of these companions are likely to be merely chance
projections of background/foreground objects and finally
derive the resulting multiplicity frequency of embedded YSOs
in Taurus and Ophiuchus (\S\,\ref{subsec:bf}).

\subsection{Visual companions to Class\,I and flat spectrum
sources} 
\label{subsec:comps}

Although embedded sources are located in dense molecular clouds,
imaging studies as deep as the one presented here can reveal a large
number of faint objects that are mostly unrelated to the star-forming
region. We therefore limited the search for companions to within a
10\arcsec\ (1400\,AU) radius around the surveyed embedded YSOs, a
cutoff which excludes the presence of many background stars projected
by chance in the vicinity of out targets (see
\S\,\ref{subsec:bf}). Using this search radius also allows to compare
directly our results to multiplicity surveys among the more evolved
T\,Tauri stars, which have been conducted over the same separation
range.  Furthermore, if the distribution of orbital periods of
protostellar multiple systems does not evolve significantly with time
(e.g., Kroupa \& Burkert 2001), we expect that this separation range
will encompass a large fraction of all binary systems, as is the case
for main sequence field stars: Duquennoy \& Mayor (1991) have for
instance shown that less than 10\,\% of solar-type field stars have a
companion at separations wider than 1400\,AU. The lower limit on
measurable binary separation is fixed by the sharpness of the images.
Systems tighter than the image's FWHM would only be slightly elongated
and could be misidentified as wind shaking or telescope tracking
errors. In order to exclude binaries that are only marginally
resolved, we do not consider systems tighter than 0\farcs8
(110\,AU). Tighter systems will be the focus of a higher resolution
study of the same sources performed with adaptive optics (Duch\^ene et
al., in prep.).

The 1400\,AU upper cutoff adopted here is several times
smaller than the size of individual prestellar cores in both
star-forming regions (Motte et al. 1998, Motte \& Andr\'e
2001). Therefore, two stars physically separated by such a
``small'' distance most likely formed through the
fragmentation of a single core. However, it has been shown
that the typical size of molecular cores differs in both
areas, with cores in Taurus being $\sim3$ times larger than
those in Ophiuchus (see \S\,\ref{sec:intro}). In order to
study the core fragmentation process, it might be more
relevant to search for binaries using a similar separation
cutoff {\it relative to the original core size}. Therefore, we
have also considered candidate companions up to 30\arcsec\
(4200\,AU) from our YSO targets in Taurus, i.e., less than a
third of the size of a typical prestellar core in that
region. In our analysis, we use the 10\arcsec\ cutoff to
define our primary sample of companions. The enlarged
separation range probed in Taurus is only used for discussing
core size-scaled binary properties in both star-forming
regions (see \S\,\ref{sec:discuss}.1).

\begin{figure*}
\includegraphics[width=0.68\textwidth]{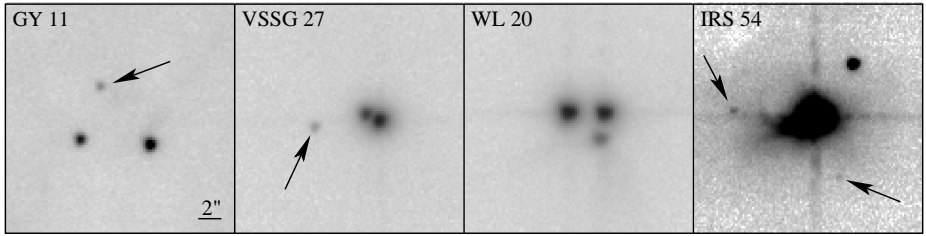} \hfill
\parbox[b]{0.3\textwidth}{\caption{Greyscale images of the candidate multiple systems detected
  in Ophiuchus. Each image is about 20\arcsec\ (2800\,AU) on a side;
  North is up and East to the left. The faintest companions are
  indicated with arrows.}}
\label{fig:triples}
\end{figure*}

In the 0\farcs8--10\arcsec\ range, we found a possible
companion to 5 of the 22 protostars we surveyed in Taurus,
while in Ophiuchus, we found 14 candidate companions
associated to 11 targets in our sample of 41 protostars. The
observed raw companion star fractions\footnote{Throughout this
study we adopt the binomial statistics for estimating
uncertainties associated with the companion star fraction,
$\sigma_{CSF} = \sqrt{N_{comp}(1 - N_{comp}/N_{prim})} /
N_{prim}$ (where $N_{prim}$ and $N_{comp}$ are the number of
primaries and companions, respectively), as a more appropriate
method for this problem than the widely used Poisson
statistics since the $N_{comp}/N_{prim}\ll 1$ criteria is
usually not met.} is 23\,$\pm$9\% in Taurus and 34\,$\pm$7\%
in Ophiuchus. In the 10--30\arcsec\ range around our Taurus
targets, we find 19 additional possible companions, although a
majority of them are likely to be projected background stars
(see \S\,\ref{subsec:bf}). Images of the possible higher-order
multiple systems with separations $\leq10$\arcsec\ are
presented in Figure~\ref{fig:triples}. The astrometric and
photometric properties of all candidate companions are
summarized in Table~\ref{tab:bin}. The faintest two companions
to IRS\,54 were detected by combining a short integration
image with a much longer exposure of the same field; they are
therefore excluded from our discussion as we obtained such
deep images for only a handful of objects in our sample.

\begin{table*}
\caption{\label{tab:bin}List of all companions detected within
10\arcsec\ (Ophiuchus) or 30\arcsec\ (Taurus) of our
targets. Flux ratios are measured with respect to the
brightest component in the $K$ band image, except for three
systems where the Class\,I source is the faintest component in
the near-infrared ($\dagger$ symbol). The boldfaced entries
represent systems that are likely to be physically bound (see
\S\,\ref{subsec:bf}).  The two companions to IRS~54 listed in
italics were detected by combining a short and a long exposure
and are not included in our survey. Upper limits for flux
ratios are at the 3\,$\sigma$ level. IRAS\,04264+2433 was
saturated in all of our images, so only a lower limit for the
flux ratio to its companion can be obtained.}
\begin{tabular}{lccccccc}
\hline
Object & $\rho$ & P.A. & $\Delta J$ & $\Delta H$ & $\Delta K$
& $\Sigma(K\leq K_B)$ & $P_{\rm bound}$ \\
 & (\arcsec) & (\degr) & (mag) & (mag) & (mag) &
 ($10^{-4}$ arcsec$^{-2}$) & \\
\hline
\hline
\multicolumn{8}{c}{Taurus ($0\farcs8<\rho<30$\arcsec)} \\
\hline
%
IRAS\,04016+2610 & 23.28 & 223.5 &  &  &
4.60$\pm$0.05 & 1.6 & 0.76 \\
IRAS\,04108+2803\,B & 22.16 & 237.1 &  &  &
0.17$\pm$0.02 & 0.4 & 0.94 \\
\hspace*{1.2cm}-- & 15.23 & 274.1 &  &  &
6.00$\pm$0.05 & 1.2 & 0.92 \\
{\bf IRAS\,04113+2758} & {\bf 4.03} & {\bf 152.9} & {\bf -0.94$\pm$0.03} &
{\bf -0.28$\pm$0.03} & {\bf 0.12$\pm$0.03} & {\bf 0.8} & {\bf 0.99} \\
\hspace*{1.2cm}-- & 24.80 & 279.8 & 2.96$\pm$0.05 & 3.54$\pm$0.05 &
3.90$\pm$0.05 & 1.6 & 0.73 \\
IRAS\,04158+2805 & 25.45 & 27.3 & -0.44$\pm$0.02 &
0.05$\pm$0.02 & 0.94$\pm$0.02 & 0.8 & 0.85 \\
\hspace*{1.2cm}-- & 28.95 & 246.5 & 3.86$\pm$0.05 & 4.23$\pm$0.05 &
5.08$\pm$0.05 & 5.3 & 0.25 \\
IRAS\,04181+2654\,B & 22.22 & 248.5 &  &  &
4.45$\pm$0.05 & 2.5 & 0.68 \\
{\bf IRAS\,04191+1523} & {\bf 6.09} & {\bf 303.7} & {\bf $>$2.05}
& {\bf 2.15$\pm$0.05} & {\bf 2.06$\pm$0.03} & {\bf 0.8} & {\bf 0.99} \\
\hspace*{1.2cm}-- & 28.30 & 290.0 & $>2.05$ & $>4.50$ &
4.80$\pm$0.10 & 6.5 & 0.20 \\
IRAS\,04239+2436 & 22.12 & 319.9 &  &  &
4.80$\pm$0.05 & 1.2 & 0.83 \\
{\bf IRAS\,04248+2612} & {\bf 4.55} & {\bf 15.1} &  &  &
{\bf 4.60$\pm$0.10} & {\bf 3.7} & {\bf 0.98} \\
IRAS\,04260+2642 & 12.30 & 279.4 &  &  &
5.67$\pm$0.10 & 9.0 & 0.65 \\
{\bf IRAS\,04263+2426} & {\bf 1.27} & {\bf 353.3} &  &  &
{\bf 1.05$\pm$0.05} & {\bf 0.4} & {\bf 0.99} \\
IRAS\,04264+2433 & 10.93 & 295.6 & $>6.1$ & $>6.4$ & $>5.9$ &
2.5 & 0.91 \\
IRAS\,04302+2247 & 22.64 & 120.5 &  &  &
3.88$\pm$0.05 & 5.3 & 0.43 \\
\hspace*{1.2cm}-- & 27.27 & 263.2 &  &  &
4.20$\pm$0.05 & 7.8 & 0.16 \\
\hspace*{1.2cm}-- & 19.34 & 114.4 &  &  &
4.40$\pm$0.08 & 8.6 & 0.37 \\
{\bf IRAS\,04325+2402} & {\bf 8.15} & {\bf 350.5} &  & {\bf
2.42$\pm$0.10} & {\bf 2.43$\pm$0.05} & {\bf 1.2} & {\bf 0.98} \\
IRAS\,04365+2535 & 12.48 & 133.6 &  &  &
6.25$\pm$0.15 & 5.7 & 0.76 \\
IRAS\,04381+2540 & 25.92 & 177.4 &  &  &
6.80$\pm$0.15 & 5.3 & 0.33 \\
{\bf IRAS\,04385+2550} & {\bf 18.90} & {\bf 342.0} & {\bf 2.73$\pm$0.03} &
{\bf 2.90$\pm$0.03} & {\bf 3.20$\pm$0.03} & {\bf 0.4} & {\bf 0.96} \\
IRAS\,04889+3042 & 12.32 & 20.5 &  &  &
4.46$\pm$0.05 & 2.4 & 0.89 \\
\hspace*{1.2cm}-- & 20.67 & 295.9 &  &  &
6.05$\pm$0.10 & 6.9 & 0.40 \\
\hline
\multicolumn{8}{c}{Ophiuchus ($0\farcs8<\rho<10$\arcsec)} \\
\hline
{\bf LFAM\,1$ \dagger$} & {\bf 9.35} & {\bf 104.9} &  &  &
{\bf -4.03$\pm$0.10} & {\bf 1.1} & {\bf 0.97} \\
{\bf GY\,11} & {\bf 6.20} & {\bf 86.1} &  &  & {\bf 0.76$\pm$0.05} &
{\bf 3.2} & {\bf 0.96} \\
\hspace*{0.4cm}-- & 6.73 & 40.9 &  &  &
2.10$\pm$0.10 & 4.5 & 0.94 \\
{\bf VSSG\,27} & {\bf 1.22} & {\bf 64.3} &  & {\bf 0.84$\pm$0.10} &
{\bf 1.32$\pm$0.05} & {\bf 0.7} & {\bf 0.99} \\
{\bf \hspace*{0.4cm}--} & {\bf 5.73} & {\bf 95.7} &  & {\bf 4.81$\pm$0.15} &
{\bf 5.29$\pm$0.05} & {\bf 2.4} & {\bf 0.98} \\ 
{\bf WL\,2} & {\bf 4.24} & {\bf 341.1} &  & {\bf 1.35$\pm$0.03} &
{\bf 1.36$\pm$0.02} & {\bf 0.4} & {\bf 0.99} \\
{\bf WL\,20$ \dagger$} & {\bf 3.66} & {\bf 51.5} &  &  &
{\bf -3.13$\pm$0.03} & {\bf 0.2} & {\bf 0.99} \\
{\bf \hspace*{0.4cm}--} & {\bf 2.24} & {\bf 351.9} &  &  &
{\bf -2.89$\pm$0.03} & {\bf 0.2} & {\bf 0.99} \\
{\bf IRS\,37 $\dagger$} & {\bf 8.68} & {\bf 63.7} &  &  &
{\bf -1.60$\pm$0.03} & {\bf 0.2} & {\bf 0.99} \\
{\bf IRS\,43} & {\bf 7.12} & {\bf 321.5} &  &  & {\bf 2.98$\pm$0.05} &
{\bf 1.3} & {\bf 0.98} \\
{\bf IRS\,51} & {\bf 1.62} & {\bf 9.0} &  &  & {\bf 3.67$\pm$0.05} &
{\bf 1.1} & {\bf 0.99} \\
{\bf IRS\,54} & {\bf 7.17} & {\bf 323.1} &  & {\bf 5.40$\pm$0.10} &
{\bf 5.10$\pm$0.03} & {\bf 1.9} & {\bf 0.97} \\
\hspace*{0.4cm}-- & \it 9.01 & \it 86.4 & \it  &
$>\mathit{5.9}$ & \it 7.97$\pm$0.05 & \it $\mathit{7.8}$ & \it 0.82 \\ 
\hspace*{0.4cm}-- & \it 7.14 & \it 202.4 & \it  &
$>\mathit{5.9}$ & \it 9.09$\pm$0.10 & \it $\mathit{2.0}$ & \it 0.73 \\ 
{\bf IRS\,67} & {\bf 2.83} & {\bf 300.0} &  &  & {\bf 4.13$\pm$0.10} &
{\bf 9.1} & {\bf 0.98} \\
ISO\,1632021-245616 & 7.78 & 179.8 &  &  &
6.7$\pm$0.2 & 15. & 0.75 \\
\hline
\end{tabular}
\end{table*}

\begin{figure}
\centering
\includegraphics[width=\columnwidth]{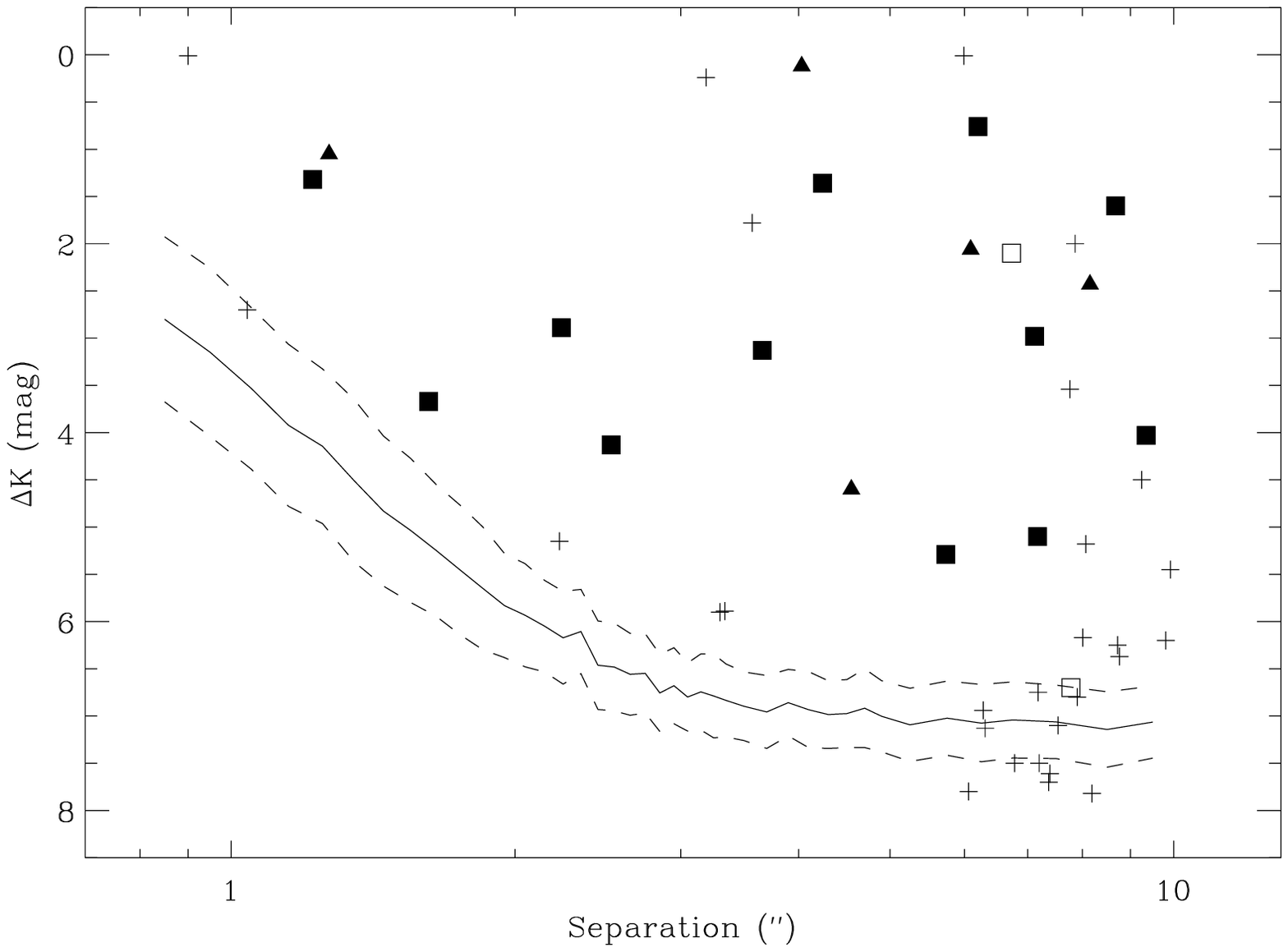}
   \caption{Observed flux ratios of our targets (Taurus: {\it
   triangles}; Ophiuchus: {\it squares}) and of
   Class\,II and III systems in Ophiuchus ({\it plus
   signs}). The filled and empty symbols for our Ophiuchus
   targets discrimate between physically bound and projected
   systems (see \S\,\ref{subsec:bf}). The solid line is our
   median 3\,$\sigma$ detection limit, and the dashed lines
   represent the 1\,$\sigma$ variation about that line (see
   text for details).}
\label{fig:fluxratios}
\end{figure}

Our ability to detect a companion depends on its flux relative
to its brighter primary: the closer the companion, the
brighter it needs to be for us to detect it. This is
illustrated in Figure~\ref{fig:fluxratios}, where we have
plotted all observed companions in our images. We have also
included in this figure companions to well-exposed Ophiuchus
T\,Tauri multiple systems that lie within our field of view
(see Appendix\,\ref{apdx}), in order to better define the
lower envelope of all detected companions. It appears that
binaries with $\Delta K\leq7$--8\,mag can be detected beyond
6\arcsec\ or so, while at 1\arcsec, only binaries with $\Delta
K\leq3$\,mag are identified. Separately, we have estimated our
ability to detect faint point sources around bright sources by
computing the standard deviation of pixel values in concentric
annuli surrounding single sources. The resulting average
3\,$\sigma$ limit is also plotted in
Figure~\ref{fig:fluxratios} along with the standard deviation
(from observations of 16 independent point sources) about that
curve. These curves match relatively well the lower envelope
of our detected companions. Within the central 2--3$\arcsec$,
the detection limit mostly depends on the seeing but it does
not appear to vary dramatically from a source to another. At
larger distances, the detection limit is mostly driven by the
background noise, which is roughly constant except if the
protostar is surrounded by some extended nebulosity.

We conclude that we have detected all possible companions that
lie above the solid curve in Figure~\ref{fig:fluxratios}. For
objects that have an extended nebulosity in our near-infrared
images, our ability to detect a nearby companion can be somewhat
impaired. As illustrated in Fig\,\ref{fig:nebul}, nebulosities
range from a faint extended halo (e.g., IRAS\,04169+2702) to
the two parallel nebulae associated with the presence of an
edge-on, optically thick circumstellar disk that blocks the
direct starlight (IRAS\,04302+2247). However, only extremely
bright nebulosities in which no point-like object can be seen,
such as IRAS\,04302+2247, significantly affect our detection
limit. Indeed, the fraction of visual companions we detected
around objects with an infrared nebulosity (9 out of 24,
38\,$\pm$10\%) is similar to that observed around point like
sources (10 out of 39, 26\,$\pm$7\%). We conclude that we
probably did not miss many companions due to the presence of
extended nebulosities over the separation range investigated
in this study.

\subsection{Comparison to previous surveys}
\label{subsec:compar}

In the Taurus-Auriga molecular cloud, this study is the first
census of multiple systems among a nearly complete sample of
embedded YSOs. We note that IRAS\,04113+2758 is a resolved
pair in 2MASS, while the binarity of IRAS\,04263+2426
($\equiv$ Haro\,6-10) has been discovered by Leinert \& Haas
(1989). The $\sim$6\arcsec\ companion to IRAS\,04191+1523 has
also been detected in the millimeter continuum with the IRAM
Plateau de Bure Interferometer, confirming that both sources
are young stellar objects with compact envelopes and/or disks
(Motte 1998). Finally, our image reveal an almost point-like
object located about 3\arcsec\ North of IRAS\,04016+2610.
Higher spatial resolution images of this system by Padgett et
al. (1999) show an extended structure, most likely to be light
scattered off the envelope or disk associated to this YSO. We
therefore reject this ``companion'' from our analysis. Of the
5 companions to Taurus embedded sources reported here, 2 are
new detections (IRAS 04248+2612, IRAS 04325+2402).

Some of the widest pairs we identified in Ophiuchus were
already known: LFAM\,1 (whose companion is GY\,12 $\equiv$
ISO\,34), IRS\,43 (its companion is GY\,263), WL\,2 (from
2MASS), WL\,20 (Ressler \& Barsony 2001), GY\,11 and IRS\,37
(Allen et al. 2002). In their recent survey of embedded YSOs,
Haisch et al. (2004) further identified the companions to
IRS\,54 and VSSG\,27. The separations and position angle
measurements for these systems agree well with their
previously published values. We further report here the
detection of companions to 2 targets that were also
investigated by Haisch et al. (2002), namely IRS\,51 and
IRS\,67. Haisch et al. (2002) detected our proposed companion
to IRS\,51 but their non-detection of the secondary at
10\,$\mu$m lead them to conclude that it was not real but
rather a patch of scattered light in a possible outflow.  The
companion appears indistinguishable from a point source at the
0\farcs6 resolution of our image of the system. We therefore
believe that the companion is real and consider that its
non-detection at 10\,$\mu$m results from the limited
signal-to-noise ratio of the image of Haisch et al. (their
Figure~1). The companion to IRS\,67 was probably not detected
by Haisch et al. because of its small separation, 2\farcs5,
and large flux ratio, $\Delta\,K\sim 4.1$\,mag. Overall, of
the 14 companions we report here for Ophiuchi embedded
sources, 3 are new detections (IRS\,51, IRS\,67 and
ISO\,1632021-245616).
 
Overall, our survey thus revealed 5 new companions in the
0\farcs8--10\arcsec\ separation range, as well as the first
infrared detection of a companion only detected at millimeter
wavelengths so far. This significantly adds to the statistics
of multiplicity of embedded YSOs in both star-forming
regions. Haisch et al. (2004) recently surveyed 6 star-forming
regions, with YSO samples ranging from 9 to 30 objects per
region, and derived an overall companion star fraction of
12/76 (16\,$\pm$4\%) over the separation range 300-1400 AU and
down to a flux ratio $\Delta$K=4.0 mag. Over the same
separation range and detection limits, we derive a companion
star fraction of 13/63 (20\,$\pm$5\%) for embedded YSOs in
Taurus and Ophiuchus. The 2 surveys yield very similar
results, even though the former includes smaller samples of
YSOs distributed over a number of star-forming regions, while
our survey deals with nearly complete YSO populations in only
2 star-forming regions. The similar results obtained from the
two surveys thus suggests that the average companion star
fraction for embedded sources does not vary much from one
star-forming region to the other. Using high resolution
radio continuum imaging, Reipurth et al. (1999, 2002, 2004)
have compiled a list of 21 embedded protostars in several
star-forming regions that includes 4 multiple systems with a
separation in the range 110--1400\,AU studied here, a
multiplicity rate also very similar to the one we find here.

\begin{figure*}
\parbox{0.6\textwidth}{\hspace*{2.5truecm}See attached JPEG figure {\tt
duchene\_f4.jpg}} \hfill
\parbox{0.4\textwidth}{\caption{Example of nebulosities surrounding some
of our targets; the upper row show some Taurus sources, the
lower Ophiuchus targets. Each image is 25\arcsec\ on the side;
North is up and East to the left. The contours are spaced by
0.75\,mag/arcsec$^2$ from the peak flux down to approximately
the 4--5\,$\sigma$ limit.}}
\label{fig:nebul}
\end{figure*}

\subsection{Physically bound multiple systems}
\label{subsec:bf}

The candidate companions we have identified could be either
physically bound to our targets or unrelated objects (YSOs or
field stars) that are projected within 10\arcsec\ (or
30\arcsec). Since we only have $K$ band photometry for most of
our candidate companions, it usually is not possible to use
color indices to determine their nature. The two visual
companions to IRAS\,04158+2805 are likely background stars,
given their location in the near-infrared color-color diagram
(see Figure~\ref{fig:jhk}). In any case, the large and
variable extinction across the molecular clouds would represent
a major issue as a much redder companion could be interpreted
both as an extincted background star or a more embedded
companion protostar.

We therefore used a statistical approach to estimate the
probability that the companions we found are physically bound
to their primary. Our method relies on the large field of view
of our images, which allows us to estimate the local surface
density of faint objects in the field. More specifically, we
used the same method as described in Duch\^ene et al. (2001),
which we summarize here briefly.

We first count the number of objects located in our field of
view that are at least as bright as the candidate companion,
including the companion itself but not its primary. We convert
this number into the average surface density of objects
brighter than this threshold, $\Sigma(K\leq K_B)$. The median
value of this quantity over all fields is about
$2\times10^{-4}$\,arcsec$^{-2}$. Assuming a random
distribution of background stars across the fields, we then
estimated the number of companions one would expect to lie as
close to the primary as the companion or closer.  This number
is $\bar{N}=\pi(\rho_B^2 - \rho_{lim}^2) \times \Sigma (K\leq
K_B)$, where $\rho_B$ is the binary separation and
$\rho_{lim}=0\farcs8$ is the smallest separation probed in
this survey.  Finally, we estimated the probability that a
random uniform distribution of stars across the field yields
no companion within $\rho_B$, which we estimate using Poisson
statistics (justified by the fact that $\bar{N}$ is almost
always smaller than unity): $P_{bound} = P(N=0) =
e^{-\bar{N}}$. The resulting probabilities of the companion
being physically associated to a YSO are given in the last
column of Table~\ref{tab:bin}. By construction, this method
takes into account the projection of background and foreground
objects, as well as of unrelated cloud members.

Among the 0\farcs8--10\arcsec\ companions detected in our
survey, 5 out of 14 in Ophiuchus and 3 out of 5 in Taurus have
probabilities of being bound larger than 99\,\%. These are
very likely physical systems but we have to decide on a
threshold down to which we accept the companions as likely.
On the other hand, most Taurus companions between 10\arcsec\
and 30\arcsec\ have probabilities lower than 90\,\% of being
bound. Those are likely to be projected unrelated stars. In
the following, we decided to consider only systems that
have probabilities of being bound larger than 95.5\,\%, i.e.,
for which a background status can be rejected at the
2\,$\sigma$ level. This is probably a conservative estimate,
as our method explicitly assumes that background stars are
randomly located in the fields while our images show that the
protostars are usually found in darker areas, with an
increasing number of sources towards the edges of our images
(i.e., YSOs are located deeper in molecular clouds). This is
however an image-to-image effect whose amplitude is difficult
to estimate due to the relatively small number of stars per
field. On the other hand, members of young associations tend
to cluster in relatively compact groups, which may produce
apparent pairs of two YSOs that are in fact unbound. For
instance, among the companions retained with this criterion,
one (GY\,12, companion of LFAM\,1) is a known member of the
Ophiuchus star-forming region (see \S\,\ref{subsec:compar})
that appears to be a Class\,III, weak-line T\,Tauri star. It is
not clear whether two objects that have such a different
evolutionary status do form a physically bound binary system
or whether this is the mere result of the projection of two
members within 10\arcsec.  In any case, background
contamination can only be estimated on a statistical basis and
there is no single threshold that perfectly separate physical
systems from projected ones. With this caveat in mind, we
adopt the 2\,$\sigma$ threshold in our analysis.
    
In our survey, we therefore have identified 5 binary systems with
separation 0\farcs$8<\rho<10$\arcsec\ out of 22 protostars in Taurus
and 12 companions to 10 of our 41 sources in Ophiuchus. This
corresponds to a companion star fraction of 23$\pm$9\,\% in Taurus,
29$\pm$7\,\% in Ophiuchus and 27$\pm$6\,\% in the 110--1400\,AU range
for the combined sample. Our detection limit is not uniform across the
separation range probed in this study and we may therefore be missing
a few close, faint companions to our targets. We do not attempt to
correct for this, as one would have to rely on assumptions regarding
the flux ratio and separation distributions of protostellar binary
systems, which are currently unknown. Consequently, the number of
companions we estimate might slightly underestimate the actual one
over the separation range investigated here. In addition, we find one
candidate companion in Taurus which is possibly physically bound,
though just at the 2$\sigma$ level, to a YSO in the range
10--30\arcsec.


\section{Discussion}
\label{sec:discuss}

The survey we have conducted yields a statistically
significant estimate of the multiplicity frequency for a
nearly complete population of embedded YSOs in Taurus and
Ophiuchus. We discuss below the implications of this result
regarding the fragmentation of pre-stellar cores as they
collapse (\S\,\ref{subsec:finalbf}). We then investigate
whether prompt dynamical evolution of multiple systems occurs
during the protostellar embedded stage, on a timescale of
$\gtrsim 10^5$\,yr, by comparing the statistics of
multiplicity we derived for Class\,I and flat spectrum sources
to that of even younger, more deeply embedded Class\,0
sources.  We further compare the frequency of embedded
multiple systems to that of Class II and III T Tauri stars in
the same clouds, in order to investigate the dynamical
evolution of multiple systems on a longer timescale of
$\gtrsim 10^6$\,yr (\S\,\ref{subsec:evol}). Finally, we
confront the predictions of various classes of models dealing
with the formation and early evolution of multiple stellar
systems to the statistics of multiplicity of embedded sources
(\S\,\ref{subsec:models}).

\subsection{Multiple protostellar systems and the fragmentation of
  collapsing cores}
\label{subsec:finalbf}

The primary result of this study is that, over the separation
range 110--1400\,AU, embedded YSOs present similar apparent
multiplicity rates in the Taurus and Ophiuchus star-forming
regions despite the known differences between these two
regions (pre-stellar core properties and stellar density).
The size of prestellar cores being $\sim3$ times larger in
Taurus than in Ophiuchus, the comparison of the properties of
binary systems spanning the same physical separation range may
introduce a bias if core fragmentation is a scale-free process
(e.g., Sterzik, Durisen \& Zinnecker 2003). We therefore also
compare the 110--1400\,AU multiplicity rate of Ophiuchus
(29$\pm$7\,\%) to that over the 330-4200\,AU separation range
in Taurus.  The latter is 23$\pm$9\,\% (5 companions out of 22
targets), statistically indistinguishable from the Ophiuchus
rate.  This alternative comparison reinforces our conclusion
that the frequency of companions among our two samples of
embedded YSOs are identical in the 2 star-forming
regions. This in turn suggests that, {\it in spite of initial
differences between Taurus and Ophiuchus pre-stellar cores,
fragmentation appears to produce very similar properties for
their multiple systems at the protostellar stage.}

Both samples of embedded YSOs we studied are nearly
complete. Nevertheless, they exhibit a notable difference:
roughly 25\,\% of the primaries in Ophiuchus have $L_{bol}\leq
0.1\,L_\odot$, while only one such source is included in the
Taurus sample. These low luminosity objects may either be
extremely low mass objects (e.g. candidate embedded brown
dwarfs) or embedded sources with reduced accretion rates. In
any event, it might not be fair to compare the multiplicity
properties of these two samples if they are made of
intrinsically different populations (different mass function
and/or evolutionary stages).  The median luminosity of both
samples is about 0.5\,$L_\odot$. We compared the multiplicity
rate of the faintest half of our samples to that of the
brightest half.  Overall, we find a marginally higher
multiplicity rate among objects with $L_{bol}>0.5\,L_\odot$
(11 companions to 33 sources) than among less luminous ones (6
out of 30), with both molecular clouds suggesting the same
trend although not at a statisticaly significant level due to
the small sample sizes. Hence, irrespective of whether we
include all embedded YSOs in our sample or not, the overall
companion star fraction over the 110--1400\,AU range is on
order of 20--30\,\% in both star-forming regions.

The observed proportion of companions to protostars we find is
relatively high, especially if one considers that hardly more
than an order of magnitude in separation is probed in this
study. In the same separation range, the companion star
fraction of field G dwarfs amount to only 14\,\% (Duquennoy \&
Mayor 1991). This comparison is relevant if the distribution
of separation for protostellar systems is as wide as that of
mature systems. Kroupa \& Burkert (2001) concluded from N-body
simulations that stellar dynamical interactions in embedded
clusters cannot significantly broaden the initial distribution
of orbital periods, so that the wide range of separations
observed for mature systems must result from the fragmentation
process itself. At an earlier stage, accretion of
material from the infalling envelope on a ``seed'' protobinary
may also change the initial separation of the binary. Through
SPH simulations, Bate (2000) has shown that this phenomenon is
largely dependent on the (unknown) initial conditions. While
we may expect accreting binary systems to get wider with time
on average, we note that the excess of close companions in the
separation range studied here is almost twice as large as the
frequency of systems with larger separations for field G
dwarfs. Hence, the excess companion star fraction we find
over a restricted range of separation in embedded sources
suggests that binary and multiple systems are indeed a very
frequent outcome of the dynamical fragmentation of prestellar
cores as they collapse.

The similitudes of the multiple systems in the 2 molecular
clouds extend beyond their frequency, and include as well
their separation distribution and $K$ band flux ratios, which
both evenly sample the ranges probed in this survey
(Figure~\ref{fig:fluxratios}). Protostellar multiple systems
appear to span a larger range of flux ratios than the more
evolved T Tauri (Class\,II) binaries, which are usually
limited to $\Delta K <3$\,mag (e.g., Leinert et
al. 1993). This may be because the observed $K$ band flux of
an embedded YSO is mostly driven by its accretion rate and
line of sight extinction, while the central source's
luminosity is usually a small fraction of the total system's
luminosity (Bontemps et al. 2001).  This is not the case for
T\,Tauri stars, in which near infrared flux ratios more
reliably trace mass ratios.  Therefore, we cannot use the
distribution of flux ratios of our sample of YSOs to infer the
underlying mass ratio distribution for these objects, but can
merely conclude that the relative accretion rates onto the
components of the multiple protostellar systems appear similar
in Taurus and in Ophiuchus, in spite of initially different
prestellar core structures.

\subsection{The evolution of protostellar multiple systems}
\label{subsec:evol}

The frequency and orbital properties of protostellar systems
is expected to evolve on various timescales due to a number of
processes, such as dynamical decay of initially unstable
configurations ($\sim$10$^{4-5}$\,yr, e.g., Delgado-Donate,
Clarke \& Bate 2003; Goodwin, Whitworth \& Ward-Thompson
2004a) or disruptive encounters with other systems if located
in dense environments ($\sim$10$^{5-6}$\,yr). In this context,
a somewhat surprising result is the similar frequency and
properties we find for protostellar systems in both Taurus and
Ophiuchus. This suggests that their early dynamical evolution
($\leq$10$^5$\,yr) has been remarkably similar, in spite of
both different initial pre-stellar core structures and
different environmental conditions, the stellar density in
Ophiuchus being 2 to 3 orders of magnitude larger than in
Taurus (e.g., Allen et al.  2002).

Searching for possible evolutionary effects during the
embedded phase, we first separate Class\,I from flat spectrum
(FS) sources within our sample, as the latter are believed to
be in an evolutionary state intermediate between Class\,I and
T\,Tauri objects, at least on a statistical basis.  Combining
the two star-forming regions in order to avoid small sample
sizes, we find 9 companions among 28 Class\,I sources and 8
companions among 35 flat spectrum primaries. This yields a
companion star fraction of 32$\pm$9\,\% for Class\,I and of
23$\pm$7\,\%, for FS sources over the 110--1400\,AU separation
range, a hardly significant evolutionary trend.


\begin{table*}
\caption{\label{tab:bfs}Observed companion star fraction over
  the separation range 110--1400\,AU in Perseus, Taurus-Auriga
  and Ophiuchus star-forming regions (SFRs) for Class\,0
  sources, Class\,I sources with (env+) and without (env-) an
  extended millimetric envelope, and for Class\,II-III T Tauri
  stars in the same clouds. For comparison, the multiplicity
  rate for G-type field dwarfs over the same separation range
  is 14\,\% (Duquennoy \& Mayor 1991). References: 1 - Looney
  et al. (2000); 2 - this study; 3 - Mathieu (1994); 4 -
  Barsony et al. (2003).}
\begin{tabular}{|c|cccc|c|c|c|}
\cline{2-7}
\multicolumn{1}{c|}{} & SFR & $N_{comp}$/$N_{prim}$ & CSF &
Ref. & CSF$_{tot}$ & CSF$_{tot}$ & \multicolumn{1}{c}{} \\
\multicolumn{1}{c|}{} & & & (\%) & & (\%) & (\%) &
\multicolumn{1}{c}{} \\
\hline
 & Perseus & 1 / 11 & & & & & \\
\raisebox{1.5ex}[0pt]{Class 0} & Ophiuchus & 2 / 2
& \raisebox{1.5ex}[0pt]{25$\pm$13}
& \raisebox{1.5ex}[0pt]{1} & \raisebox{1.5ex}[0pt]{25$\pm$13} &
& \raisebox{0.25ex}[0pt]{Embedded sources} \\ 
\cline{1-6}
 & Taurus & 5 / 12 & 41$\pm$14 & & &
\raisebox{1.5ex}[0pt]{{\bf 38$\pm$8}} & \\
\raisebox{1.5ex}[0pt]{Class I env+} & Ophiuchus & 7 / 15 &
47$\pm$13 & \raisebox{1.5ex}[0pt]{2} &
\raisebox{1.5ex}[0pt]{44$\pm$10} & & \raisebox{2.75ex}[0pt]{with
mm envelope} \\ 
\hline
\hline
 & Taurus & 0 / 10 & 0 & & & & \\
\raisebox{1.5ex}[0pt]{Class I env-} & Ophiuchus & 5 / 26 & 19$\pm$8 &
\raisebox{1.5ex}[0pt]{2} & \raisebox{1.5ex}[0pt]{14$\pm$6} & &
\raisebox{0.25ex}[0pt]{Optical/IR sources} \\
\cline{1-6}
 & Taurus & 24 / 100 & 24$\pm$4 & 3 & &
\raisebox{1.5ex}[0pt]{{\bf 22$\pm$3}} & \\
\raisebox{1.5ex}[0pt]{Class II-III} & Ophiuchus & 21 / 88 & 24$\pm$5 & 4
& \raisebox{1.5ex}[0pt]{24$\pm$3} & &
\raisebox{2.75ex}[0pt]{without mm envelope} \\ 
\hline
\end{tabular}
\end{table*}

Our source selection is, however, exclusively based on the
near- to mid- infrared spectral index, which may not be the
most appropriate tracer of a YSO's evolutionary status. For
instance, T\,Tauri stars that are observed with an optically
thick edge-on disk (such as IRAS\,04158+2805 in our sample)
can have a spectral index typical of a Class\,I source. A more
reliable indicator of the evolutionary status of an embedded
YSO is the presence of an extended envelope, which
characterizes a {\it bona fide} protostar.  Such a
protostellar envelope can be directly traced by its thermal
millimeter continuum emission (e.g. Andr\'e \& Montmerle 1994,
Motte \& Andr\'e 2001) and often indirectly traced by an
extended near-IR nebulosity (Park \& Kenyon 2002 and this
study).  The presence/absence of such an envelope is listed
for all sources in our sample in Tables\,\ref{tab:sample1} and
\ref{tab:sample2}. Twelve out of 22 YSOs in Taurus and 15 out
of 41 YSOs of Ophiuchus exhibit evidence for a protostellar
envelope. Combining the 2 regions, we find 12 companions to
the 27 true protostars surveyed giving a companion star
fraction of 44$\pm$10\,\% in the 110--1400\,AU range (see
Table~\ref{tab:bfs}).  In contrast, the more evolved sources
in our sample, which have already dissipated their envelope,
have a much lower companion star fraction: 14$\pm$6\,\% (see
Table~\ref{tab:bfs}).  Hence, adopting the presence of an
envelope as a qualitative diagnostic of youth, we find that
young self-embedded sources tend to be more often in multiple
systems than more evolved ones. {\it This evolution of the
multiplicity rate, seen here at the 2.6$\sigma$ level,
suggests a significant reduction of the number of wide
multiple systems over a timescale of $\sim 10^5$\,yr}, which
could conceivably result either from disruptive encounters
with other systems or from the internal decay of initially
unstable systems.

If this trend can be extrapolated back to even earlier stages
of evolution, one may expect to find an even larger fraction
of multiple systems among Class\,0 sources.  The investigation
of 14 YSOs driving large scale outflows led Reipurth (2000) to
claim a multiplicity frequency in the range 79--86\%, much
higher than found here for Class\,I and flat spectrum
sources. However, Reipurth's (2000) sample contains sources
that span the full range from Class\,0 protostars to T Tauri
stars and some of the multiple systems have separations
outside the range studied here. Moreover, if giant Herbig-Haro
flows are actually triggered by the decay of unstable
protostellar multiple systems, as Reipurth (2000) argues, this
sample is obviously strongly biased towards the detection of
such systems. Therefore, the derived multiplicity frequency of
$\sim 80\%$ probably does not represent the true fraction of
multiple systems in deeply embedded YSOs.  A more meaningful
comparison can be made with the results of the 2.7\,mm
continuum interferometric survey of embedded YSOs performed by
Looney, Mundy \& Welch (2000) with a spatial resolution
similar to ours. Their sample contains 12 ``separate
envelope'' Class\,0 sources (see their Table 5, assuming that
the members of the SVS13 group are all Class\,0 objects and
classifying SSV13B as a separate envelope source), i.e.,
sources which would be considered as distinct primaries in our
study.  Among these, they find 3 companions in the separation
range 110--1400\,AU corresponding to their definition of
``common envelope systems'', to which their millimeter
interferometric observations are mostly sensitive. In this
range, the companion star fraction of their Class\,0 sample is
thus 3/12 = 25$\pm$13\,\%, including 2 companions to 2 sources
in Ophiuchus. Such a fraction is consistent with the
multiplicity rate we derive for Taurus Class\,I sources with
extended envelopes (see Table~\ref{tab:bfs}).  Unfortunately,
the poor statistics stemming from the small sizes of the
samples and the differing separation sensitivities of the
millimeter interferometric survey and the present near-IR
infrared survey prevent us from drawing any strong conclusion
about the evolution of the multiplicity rate between Class\,0
and Class\,I sources on a timescale $<$10$^5$\,yr.

Looking further downstream in the evolution of young multiple
systems, we can also compare the multiplicity frequency we
derived for embedded sources in Taurus and Ophiuchus to the
fraction of multiple systems found among more evolved
Class\,II and Class\,III T Tauri stars in the same regions.
For the latter groups, we used the lists of multiple systems
from Mathieu (1994) in Taurus and Barsony, Koresko \& Matthews
(2003) in Ophiuchus, which likely contain most
0\farcs8--10\arcsec\ companions, as those can be found from
direct infrared imaging. Beyond 1\arcsec, those surveys are
likely to be complete and they actually contain systems
with flux ratios as large as 4\,mag at least, thus similar to
our detection limits. In the separation range considered here,
there are 21 companions to 88 optically revealed T Tauri stars
in Ophiuchus and 24 companions to 100 primaries in Taurus (see
Figure~\ref{tab:bfs}). The two populations have a companion
star fraction of 24$\pm$3\,\%, i.e., a factor of about 2 lower
than that of Class\,I sources with envelopes
(44$\pm$10\,\%). The much lower multiplicity fraction of
Class\,II-III sources in the 2 clouds compared to that of the
youngest Class\,I sources surrounded by millimeter envelopes (a
1.9$\sigma$ difference) seems to confirm the similar trend
reported above between young and evolved embedded Class\,I
sources.

Table~\ref{tab:bfs} summarizes the statistics of multiplicity among
embedded sources and young PMS stars in Taurus and Ophiuchus.  The
most significant result, yet still relying on relatively small
samples, is provided by the comparison between deeply embedded
Class\,0 and Class\,I sources surrounded by extended millimeter
envelopes on the one hand, and more evolved, envelope-less
Class\,I-II-III sources, on the other. The companion star fraction is
seen to decrease from 38$\pm$8\,\% in the youngest embedded sources
(age$\lesssim 10^5$\,yr) down to 22$\pm$3\,\% in pre-main sequence
objects (age $\sim$10$^6$\,yr). Overall, we thus find evidence for an
evolution of the multiplicity fraction during the embedded phase, with
a decrease of multiple systems over a timescale of $10^6$\,yr at most,
significant at the 2$\sigma$ level. The comparison between Class\,I
sources with and without extended envelope suggests that the actual
timescale is on the order of $\sim 10^5$\,yr.

We caution, however, that this result is limited by relatively small
number statistics and strongly depends on the reliability of the
evolutionary classification adopted for the YSOs in our sample. While
the mass and/or optical depth of the envelope is generally recognized
as being a robust tracer of the circumstellar evolution of embedded
YSOs (e.g. Myers et al. 1998; Andr\'e, Ward-Thompson, \& Barsony
2000), it is more difficult to obtain reliable estimates of envelope
properties in cluster-forming regions such as the Ophiuchus main cloud
than in regions of `isolated' star formation such as Taurus. In
Taurus, the relatively large spatial extent of prestellar cores and
protostellar envelopes ($\sim 18\,000$~AU) makes it easy to
distinguish between bona-fide protostars with envelopes and more
evolved Class\,I sources (Motte \& Andr\'e 2001).  In Ophiuchus,
however, the smaller prestellar fragmentation lengthscale leads to
more compact protostellar envelopes which are more difficult to
spatially resolve with single-dish millimeter telescopes. The current
classification of Ophiuchus Class\,I sources in objects with or
without extended envelopes (see Table\,\ref{tab:sample2}, Andr\'e \&
Montmerle 1994; Motte et al. 1998) is thus more uncertain than in
Taurus, pending the results of interferometric measurements in
progress. If we consider the Taurus cloud alone, the decrease in the
multiplicity fraction from self-embedded protostars to envelope-less
objects becomes apparently stronger, although the sample size is then
only barely statistically significant (see Table~\ref{tab:bfs}). In
this framework, our results suggest a steep decrease of the fraction
of multiple systems occurring during the protostellar embedded phase
as the envelope is being dissipated on a timescale of
$\sim$10$^5$\,yr.

\subsection{Confrontation with model predictions}
\label{subsec:models}

The above results indicate that the fraction of wide (110-1400\, AU)
protostellar multiple systems is initially high and seems to decrease
on a timescale of $\sim 10^5$\,yr during the Class\,I phase in the
Taurus and Ophiuchus clouds. How do these results compare with models
of the formation and evolution of multiple systems ?

Assuming that the distribution of orbital periods of embedded
multiple systems is at least roughly as wide as that of T
Tauri and field binaries (Kroupa \& Burkert 2001), the
multiplicity fraction we derive over a restricted range of
separation suggests that, on average, each embedded YSO has at
least one companion. This result is consistent with a class of
models which rely on the assumption that all stars are
initially in binary systems and investigate their subsequent
dynamical evolution (e.g., Larson 1972; Kroupa 1995b). Such
models predict that the initially high fraction of multiple
protostellar systems significantly decreases on a timescale of
1\,Myr or less in dense protostellar clusters, such as ONC,
due to frequent disruptive encounters between systems (Kroupa,
Aarseth \& Hurley 2001), whereas the dynamical evolution is
much milder in loose star-forming regions, such as Taurus,
where gravitational encounters are rarer.

The high fraction of embedded multiple systems we report here
for Taurus is consistent with the predictions of these models
for low density regions.  The small amount of dynamical
evolution expected to occur in Taurus will take place very
early on ($\leq 10^5$\,yr), when the initial protostellar
aggregates still have a density of order of 10$^3$\,stars
pc$^{-3}$ (Kroupa \& Bouvier 2003).  Disruptive encounters
between protostellar systems at this stage may result in the
decreasing multiplicity fraction we observe in Taurus during
the Class\,I phase on a timescale of $\sim 10^5$\,yr
(Table~\ref{tab:bfs}). Past this early stage, the stellar
aggregates quickly expand on a timescale of $\sim 10^6$\,yr to
reach the currently observed densities of order of
10-100\,stars pc$^{-3}$ and the Taurus population is not
expected to dynamically evolve any more. Thus, the higher
multiplicity rate of both embedded YSOs and T Tauri stars in
Taurus with respect to field stars suggests that loose
star-forming regions are not the dominant mode of star
formation, i.e., that only a small proportion of all field
stars form in such areas.

On the other hand, Ophiuchus is a young embedded cluster
similar to the ``dominant-mode cluster'' of Kroupa's (1995a)
models, from which most field stars are expected to arise
(Clarke, Bonnell \& Hillenbrand 2000; Carpenter 2000; Adams \&
Myers 2001). In Ophiuchus, as for Taurus, the initial decrease
of multiplicity fraction we observe during the Class\,I
embedded phase may be accounted for by disruptive encounters
occurring in high density protostellar subclusters on a
timescale of $\sim$10$^5$\,yr (Bonnell, Bate \& Vine
2003). Unlike Taurus, however, the remaining excess of
binaries observed among Class\,II-III T\,Tauri stars in
Ophiuchus at an age of $\sim$10$^6$\,yr has still to be
reduced by a factor of 2 to be consistent with the frequency
of field binaries. The current average density of about
10$^3$\,stars pc$^{-3}$ in the Ophiuchus cloud translates into
a crossing time of order of 1--2\,Myr (e.g., Kroupa 1995a),
which may indicate that the T\,Tauri population is not fully
dynamically evolved yet in this cloud.

Together with observations of embedded multiple YSOs in an
even denser cluster such as Orion, N-body simulations of the
evolution of an Ophiuchus-like cluster would provide a crucial
test on the assumption that core fragmentation produces a
universal set of initial properties for protostellar multiple
systems which subsequently evolve depending only upon
environmental conditions. We note however that the similar
frequency, properties and evolution we derive for embedded
multiple systems in both Taurus and Ophiuchus seems to argue
against large-scale environmental conditions playing a crucial
role in shaping the properties of protostellar systems.
Instead, the similar evolution of protostellar multiple
systems in star-forming regions whose {\it average} properties
(e.g., mean stellar density) are so different may point to
local conditions prevailing in e.g.  small-scale embedded
subclusters as being determinant (Bonnell et al. 2003;
Sterzik et al. 2003).

The observed high frequency of multiple systems among embedded YSOs
indicates that most prestellar cores undergo fragmentation into at
least two objects during their collapse. It is often assumed that
prestellar cores actually experience multiple fragmentation as they
collapse thus leading to the formation of several protostellar seeds,
i.e., small-N protostellar aggregates with $N = 3$--10, typically, on
a scale of a few 100\,AU (e.g.  Burkert, Bate \& Bodenheimer 1997;
Sterzik \& Durisen 1998; Delgado-Donate et al. 2003; Goodwin et
al. 2004a). Global simulations of cloud collapse support this scenario
and predict the formation of even higher density subclusters
consisting of 10-50 protostellar seeds (Klessen, Burkert \& Bate 1998;
Bate, Bonnell \& Bromm 2003; Bonnell et al. 2003). Small-N
protostellar aggregates as well as high density subclusters undergo
rapid dynamical evolution on a timescale of 10$^{4-5}$\,yr (Sterzik \&
Durisen 1998; Delgado-Donate et al.  2003). The dynamical decay of
initially unstable groups is thus expected to eject most of the lowest
mass protostellar seeds, with a typical ejection velocity of
$\lesssim$ 1 km/s, while those remaining in the cloud core form stable
binary and higher order multiple systems.

At an age of about 10$^5$\,yr, we find no evidence that
embedded YSOs in Taurus and Ophiuchus belong to small-N
clusters or high density subclusters on a scale of a few
100\,AU. No triple embedded YSO is found in Taurus, and only 2
in Ophiuchus, compared to a total of 13 binary YSOs detected
over the separation range 110--1400\,AU. It may be argued that
dynamical evolution of small-N clusters has already taken
place by that time, ejecting most of the initial protostellar
seeds at larger distances ($\gg$1000 AU) and leaving only the
resulting stable systems. However, Allen et al. (2002)
investigated the spatial distribution of prestellar cores and
YSOs in Ophiuchus and concluded that the YSOs had neither
concentrated nor dispersed significantly since their formation
in these cloud cores. Numerical simulations of the dynamical
decay of small-N clusters including accretion onto the
protostellar seeds and circumstellar disks
(e.g. Delgado-Donate et al. 2003, Goodwin et al.  2004a)
predict the formation of an equal number of tight binaries and
higher order (triple and quadruple) systems. This prediction
also seems to contrast with our results, although it may
simply indicate that a fraction of our sample primaries
actually are tight binaries with a separation less than
100\,AU and have therefore gone unresolved in this
study. Pending a higher resolution study of these embedded
sources (Duch\^ene et al., in prep.), a comparison of the
hierarchy of multiplicity as derived from observations to
model predictions remains somewhat inconclusive.

For the time being, on the sole basis of the high frequency of
wide binaries among embedded Class\,I and flat spectrum
sources, we cannot firmly distinguish between the formation of
unstable small-N aggregates and a core fragmentation process
that only creates 2 or 3 seeds, i.e.  stable multiple systems
without further internal dynamical evolution. In any event,
the similar multiplicity rate we derive for embedded YSOs in
both Taurus and Ophiuchus in spite of initially different
prestellar core sizes and widely different average stellar
density between the 2 star-forming regions indicates that,
regardless of the formation channel, the outcome is fairly
insensitive to both initial and environmental conditions in
this range.  Whether this is accounted for by current models
remains to be seen (e.g., Delgado-Donate, Clarke \& Bate 2004;
Goodwin, Whitworth, \& Ward-Thompson 2004b; Sterzik et
al. 2003).


\section{Conclusions}
\label{sec:concl}

We have conducted a survey of visual companions to 63 embedded
YSOs in the Taurus-Auriga and Ophiuchus star-forming
regions. The main results are as follows:
\begin{itemize}
\item The companion star fraction of Class I and flat spectrum
  embedded sources over a separation range 110\,-1400\,AU is
  23$\pm$9\,\% in Taurus, 29$\pm$7\,\% in Ophiuchus and
  27$\pm$6\,\% for the combined sample. This fraction is twice
  as large as the frequency of companions to G dwarfs in the
  solar neighborhood, in the same separation range.  Assuming
  that the distribution of orbital periods of protostellar
  multiple systems is about as wide as that of mature systems,
  this high multiplicity fraction suggests that each embedded
  YSO has, on average, at least one companion.  The
  fragmentation of prestellar cores as they collapse thus
  appears to be a frequent occurrence indeed.
\item In spite of the known differences in the properties of
  prestellar cores in Taurus and Ophiuchus and the widely
  different stellar density in the 2 star-forming regions, we
  find identical properties for their embedded multiple
  systems (frequency, separations, luminosity ratios).  This
  suggests that the outcome of the fragmentation of prestellar
  cores is fairly insensitive to both initial and
  environmental conditions, at least for the range of physical
  conditions encountered in these 2 clouds.
\item Combining the Taurus and Ophiuchus clouds, the fraction
  of multiple systems appears to decrease on a timescale of
  order of $\sim 10^5$\,yr during the embedded phase, with
  the youngest embedded sources, Class\,0 and Class\,I YSOs
  still surrounded by extended millimetric envelopes,
  exhibiting nearly twice as many multiple systems as more
  evolved Class\,I YSOs devoid of mm envelopes and
  Class\,II-III T Tauri stars. This suggests that rapid disruption
  of wide protostellar multiple systems occurs during the
  embedded phase, resulting either from the internal decay of
  initially unstable systems or from disruptive encounters
  between stable systems in high density protostellar
  aggregates.
\item Over a scale of a few 100\,AU, embedded YSOs at an age $\sim
  10^5$\,yr in Taurus and Ophiuchus clouds are found to be in relative
  isolation and not assembled in subclusters. This seems to contrast
  with the assumption that protostellar collapse leads to small-N
  clusters (N=3-10) as well as with global simulations of cloud
  collapse which predict the formation of even higher density
  subclusters on these scales.  Evidence for such clustering
  properties may have to be searched for on either smaller scales
  (tight binaries) or larger ones (ejected protostellar seeds) if
  dynamical evolution has already occurred.
\end{itemize}
 
Taurus and Ophiuchus clouds probe different modes of star
formation, the former being a low density association
harboring only low mass objects, while the latter is a dense
cluster including at least one massive B star as well as a
large population of low mass T Tauri stars. The results
reported here unexpectedly suggest that the properties and
early evolution of embedded multiple systems in these two
clouds are quite similar and therefore do not seem to depend
much on initial and/or global environmental conditions.  This
may indicate that the properties of multiple systems are
ultimately driven by dynamical processes acting on small
scales and are thus shaped by local (rather than global)
conditions which might not be very different between clusters
and associations during the early stages of protostellar
evolution.


\begin{acknowledgements}
  We are grateful to Bo Reipurth and to an anonymous referee for
  enriching comments that helped us clarify this manuscript. We extend
  our thanks to the CFHT and ESO/NTT staffs who supported us very
  efficiently during our observing runs. This study has made use of
  the SIMBAD database, operated at CDS, Strasbourg, France, and of the
  Two Micron All Sky Survey, which is a joint project of the
  University of Massachussetts and the Infrared Processing and
  Analysis Center, funded by the National Aeronautics and Space
  Administration and the National Science Foundation. Partial funding
  from the European Research Training Network on ``The Formation and
  Evolution of Young Stellar Clusters'' (HPRN-CT-2000-00155) is
  ackowledged.
\end{acknowledgements}



\appendix
\section{T\,Tauri stars in Ophiuchus}
\label{apdx}

In the course of our survey, we observed many objects that are
members of the Ophiuchus star-forming regions but are in the
more evolved classical and weak-line T\,Tauri (Class\,II and
III respectively) stages. For completeness, we list here all
the objects we observed (Table~\ref{tab:apdx_sample}) as well
as the candidate companions detected within 10\arcsec\ of them
(Table~\ref{tab:apdx_comps}).  Among the companions listed in
Table~\ref{tab:apdx_comps}, only 3 were previously known,
namely the ISO\,60--61 pair (listed in Bontemps et al. 2001)
and the companions to ISO\,95 and ISO\,204, listed in Haisch
et al. (2002).

It can be noted that we detected a much larger fraction of
{\it candidate} companions among this sample of T\,Tauri stars
than among the protostars that form the primary sample of this
study (51$\pm$6\,\% vs. 30$\pm$6\,\%). This is most likely a
consequence of the fact that the protostars are located deeper
in the molecular cloud and therefore they are surrounded by
comparatively darker areas in our images. On the other hand,
most T\,Tauri stars are located in areas that probably
contain many background stars that are not completely
extincted, which leads to a larger number of projected visual
pairs. This is confirmed by the generally large flux ratios
between the T\,Tauri stars and their apparent companions:
almost 75\,\% of all pairs listed in
Table~\ref{tab:apdx_comps} have $\Delta K \geq 5$\,mag while
only 3 such pairs are found in Table~\ref{tab:bin}.

\begin{table*}
\caption{\label{tab:apdx_sample}List of all T\,Tauri stars
(Class II and III objects) in Ophiuchus that were detected in
the vicinity of our target protostars. Objects whose name
starts with ``B'' are from Barsony et al. (1997). The
classification is determined from the infrared spectral
indices following Bontemps et al. (2001).}
\begin{tabular}{clc|clc|clc}
\hline
ISO\,\# & Object & Class & ISO\,\# & Object & Class & ISO\,\#
& Object & Class \\
\hline
9 & SKS\,1--4 & II & 76 & GY\,146 & II & 136 & GY\,258 & III
\\ 
13 & B\,162607-242725 & II & 78 & VSSG\,5/GY\,153 & II & 140 &
GY\,262 & II \\
14 & B\,162607-242742 & III & 79 & GY\,154 & II & 142 &
VSSG\,25/GY\,267 & II \\
19 & GSS\,29 & II & 81 & VSSG\,7/GY\,157 & III & 144 &
IRS\,45/GY\,273 & II \\
22 & B\,162618-241712 & III & 82 & GY\,163 & III & 150 &
ISO\,1627309-242734 & II \\
23 & SKS\,1--10 & II & 84 & WL\,21/GY\,164 & II & 154 &
GY\,291 & II \\
24 & VSSG\,1 & II & 86 & IRS\,26/GY\,171 & II & 155 & GY\,292
& II \\
25 & CRBR\,17 & III & 87 & B\,162658-241836 & II & 156 &
GY\,295 & III \\
28 & B\,162621-241544 & III & 89 & WL\,14/GY\,172 & II & 164 &
GY\,310 & II \\
30 & GY\,5 & II & 90 & WL\,22/GY\,174 & II & 166 & GY\,314 &
II \\
35 & GY\,215 & II & 91 & VSSG\,8/GY\,181 & III & 171 & GY\,323
& II \\
37 & LFAM\,3/GY\,21 & II & 92 & WL\,16/GY\,182 & II & 172 &
GY\,326 & II \\
39 & S\,2/GY\,23 & II & 93 & GY\,188 & II & 173 &
IRS\,53/GY\,334 & III \\
40 & El\,24 & II & 94 & B\,162703-242007 & II & 176 & GY\,350
& II \\
41 & GY\,29 & II & 95 & WL\,1/GY\,192 & II & 177 & GY\,352 &
II \\
42 & VSSG\,29/GY\,37 & III & 98 & GY\,195 & II & 179 & GY\,370 &
III \\
48 & S\,1/GY\,70 & III & 100 & B\,162705-244013 & III &
183 & GY\,377 & III \\
51 & B\,162636-241554 & II & 101 & IRS\,30/GY\,203 & III & 189 &
GY\,412 & III \\
52 & VSSG\,4/GY\,81 & II & 104 & GY\,207 & III & 190 &
GY\,450 & II \\
57 & B\,162641-241801 & III & 107 & GY\,213 & II & 191 & GY\,463 &
III \\
58 & WL\,8/GY\,96 & III & 114 & WL\,19/GY\,227 & II/III & 192 &
GY\,472 & III \\
60 & GY\,101 & III & 120 & IRS\,34/GY\,239 & II & 201 &
ISO\,1631487-245432 & II \\
64 & VSSG\,11 & III & 122 & IRS\,36/GY\,241 & II & 204 &
L1689--IRS\,5 & II \\
71 & GY\,130 & III & 125 & WL\,5/GY\,246 & III & 205 &
ISO\,1631531-245504 & II \\
72 & WL\,18/GY\,129 & II & 126 & GY\,248 & III & 208 &
ISO\,1631591-245442 & II \\
74 & IRS\,20/GY\,143 & III & 130 & SR\,12/GY\,250 & III \\
\hline
\end{tabular}
\end{table*}

\begin{table*}
\caption{\label{tab:apdx_comps}List of all companions to
T\,Tauri stars in Ophiuchus that were detected during the
course of our survey.}
\begin{tabular}{cccccc|cccccc}
\hline
ISO\,\# & $\rho$ & P.A. & $\Delta H$ & $\Delta K$ & note &
ISO\,\# & $\rho$ & P.A. & $\Delta H$ & $\Delta K$ & note \\
 & (\arcsec) & (\degr) & (mag) & (mag) & &  & (\arcsec) &
(\degr) & (mag) & (mag) & \\
\hline
9 & 7.18 & 240.0 &  & 6.74$\pm$0.10 & & 95 & 0.87 &
322.8 & 0.10$\pm$0.10 & 0.01$\pm$0.03 & \\
23 & 9.92 & 154.4 &  & 5.45$\pm$0.05 & & 126 & 5.60 &
214.7 &  & $>6.0$ & b \\
-- & 7.54 & 118.8 &  & 7.1$\pm$0.2 & & 130 & 8.76 &
162.9 &  & 6.37$\pm$0.05 & \\
24 & 6.68 & 102.4 &  & 8.51$\pm$0.10 & & 142 & 8.72 &
312.5 & $>5.2$ & 6.25$\pm$0.10 & \\
-- & 7.64 & 131.1 &  & 10.1$\pm$0.2 & & 154 & 7.86 &
196.5 & 2.7$\pm$0.2 & 2.00$\pm$0.03 & \\
30 & 8.19 & 19.9 &  & 7.82$\pm$0.10 & & 156 & 6.31 &
84.1 &  & 7.13$\pm$0.10 & \\
35 & 7.37 & 282.4 &  & 7.7$\pm$0.2 & & 166 & 3.30 &
102.2 &  & 5.9$\pm$0.2 & \\
41 & 7.76 & 273.1 &  & 3.54$\pm$0.10 & & 172 & 9.81 &
254.0 &  & 6.20$\pm$0.05 & \\
42 & 7.90 & 341.0 &  & 6.80$\pm$0.15 & & -- & 6.44 &
100.4 &  & 7.48$\pm$0.10 & \\
51 & 6.06 & 56.6 &  & 7.8$\pm$0.2 & & -- & 8.07 & 80.4
&  & 5.18$\pm$0.05 & \\
-- & 6.28 & 104.0 &  & 6.94$\pm$0.10 & & 176 & 2.23 &
161.4 & 6.6$\pm$0.2 & 5.15$\pm$0.05 & \\
52 & 7.39 & 90.7 & $>5.9$ & 7.61$\pm$0.15 & & -- & 3.34 & 55.8
& 7.1$\pm$0.2 & 5.89$\pm$0.05 & \\
57 & 1.51 & 241.6 &  & 1.47$\pm$0.10 & & 183 & 6.21 &
256.2 &  & 6.68$\pm$0.05 & \\
60 & 5.99 & 348.4 & -0.85$\pm$0.03 & 0.01$\pm$0.02 & a & -- &
8.71 & 261.2 &  & 7.6$\pm$0.2 & \\
64 & 4.9 & 134 &  & $>7.2$ & b & 189 & 6.78 & 115.9
&  & 7.5$\pm$0.2 & \\
72 & 3.57 & 291.1 &  & 1.78$\pm$0.05 & & -- & 7.20 &
113.0 &  & 7.5$\pm$0.2 & \\
74 & 1.04 & 206.6 &  & 2.7$\pm$0.2 & & 190 & 8.01 &
84.4 &  & 6.17$\pm$0.15 & \\
81 & 7.97 & 186.8 &  & 6.49$\pm$0.10 & & 204 & 2.95 &
241.4 &  & 0.30$\pm$0.10 & c \\
82 & 9.25 & 204.5 & $>2.7$ & 4.50$\pm$0.10 & & 205 & 6.39 &
318.8 &  & 6.2$\pm$0.1 & \\
86 & 7.28 & 78.3 &  & 7.8$\pm$0.2 & \\
\hline
\multicolumn{12}{l}{Notes -- a: The companion to this source
is ISO\,61 ($\equiv$GY\,103), a Class\,III object.}\\
\multicolumn{12}{l}{b: The primary is saturated at $K$ band;
no observation is available in $H$ band.}\\
\multicolumn{12}{l}{c: The primary is slightly saturated.}\\
\end{tabular}
\end{table*}

\end{document}